\pdfoutput=1

\documentclass[
    ,final            
  ]
  {aipproc}

\layoutstyle{6x9}

\usepackage{subfigure}
\usepackage{bm}

\begin{document}

\title{Soft Classification of Diffractive Interactions at the LHC}


\author{Mikael Kuusela}{
  address={Division of Elementary Particle Physics, Department of Physics, PO Box 64, FI-00014 University of Helsinki, Finland},
  altaddress={Department of Information and Computer Science, Aalto University School of Science and Technology, PO Box 15400, FI-00076 Aalto, Finland}
}

\author{Eric Malmi}{
  address={Division of Elementary Particle Physics, Department of Physics, PO Box 64, FI-00014 University of Helsinki, Finland},
  altaddress={Department of Information and Computer Science, Aalto University School of Science and Technology, PO Box 15400, FI-00076 Aalto, Finland}
}

\author{Risto Orava}{
  address={Helsinki Institute of Physics, PO Box 64, FI-00014 University of Helsinki, Finland},
  altaddress={Division of Elementary Particle Physics, Department of Physics, PO Box 64, FI-00014 University of Helsinki, Finland}
}

\author{Tommi Vatanen}{
  address={Helsinki Institute of Physics, PO Box 64, FI-00014 University of Helsinki, Finland},
  altaddress={Department of Information and Computer Science, Aalto University School of Science and Technology, PO Box 15400, FI-00076 Aalto, Finland}
}

\begin{abstract}
Multivariate machine learning techniques provide an alternative to the rapidity gap method for event-by-event identification and classification of diffraction in hadron-hadron collisions. Traditionally, such methods assign each event exclusively to a single class producing classification errors in overlap regions of data space. As an alternative to this so called hard classification approach, we propose estimating posterior probabilities of each diffractive class and using these estimates to weigh event contributions to physical observables. It is shown with a Monte Carlo study that such a soft classification scheme is able to reproduce observables such as multiplicity distributions and relative event rates with a much higher accuracy than hard classification.
\end{abstract}

\maketitle


\section{Introduction}

Diffraction is usually identified based on large rapidity gaps (LRG) although it is widely acknowledged that this requirement alone leads to insufficient separation between diffractive and non-diffractive events. This is due to long range correlations that may destroy the LRG. In fact, the gap survival probability $S^2$ of single diffractive events at LHC energies is only of the order of 10\% \cite{khoze2006}. Additionally, because of fluctuations in hadronization, also the non-diffractive background contains a non-negligible amount of LRG events \cite{khoze2010}. Moreover, a rapidity gap may just be an experimental artifact due to high detection thresholds. Hence, in order to achieve more efficient identification of diffraction, alternatives to a simple cut on $\Delta \eta$ should be investigated.

In this paper, we study the use of multivariate classification algorithms for identification of diffraction and also discriminating between single diffractive and double diffractive events. Such an approach does not explicitly look for rapidity gaps, but instead considers the full event topology in an optimal manner. That is, instead of heuristically determining the type of events to look for, such algorithms are able to learn the event characteristics providing the best discriminative power based on a suitably selected training set of labeled events.

Most classification algorithms perform a mapping of each observation into a single class. We call these \emph{hard classification} algorithms, examples of which are neural networks and support vector machines. In our case, hard classification corresponds to classifying each event as either single diffractive, with the diffractive system on the left (SDL) or the right side (SDR), double diffractive (DD) or non-diffractive (ND)\footnote{See \cite{kuusela2010} for a feasibility study of such a classification scheme.}. As there is inherent mixing between these classes, such an approach is bound to produce classification errors in the overlap regions of the data space. This is especially the case with DD events which often exhibit characteristics similar to SD and ND events. For this reason, instead of considering a single class only, we propose estimating the probabilities for each event to belong to each of the classes. We then use these probabilities to weigh the contribution of an event to physical observables. In the spirit of \cite{wahba1993}, we call such an approach \emph{soft classification}.

\section{Soft Classification Methodology}

In this work, we estimate the posterior probability of an event $\bm{x}$ to belong to class $C_i$ using the $k$ nearest neighbors ($k$NN) algorithm\footnote{We also experimented with more advanced soft classification methods such as kernel density estimation and non-linear discriminant analysis but they gave no advantage over $k$NN.} for which $p(C_i|\bm{x}) = k_i/k$, where $k_i$ is the number of observations from class $C_i$ among the $k$ nearest neighbors of $\bm{x}$ in the training set \cite{alpaydin2010}. The nearest neighbors are found using the Euclidean distance although other distance metrics can be used as well. In addition to soft classification, $k$NN can also be used for hard classification in which case the class is selected based on the highest posterior probability.


Because of an effect known as \emph{curse of dimensionality}, the performance of the $k$NN algorithm can be significantly improved by reducing the dimensionality of the data. To this end, we use the linear discriminant analysis (LDA) algorithm which is a dimensionality reduction algorithm for labeled data \cite{alpaydin2010}. It performs a mapping $\bm{x} \mapsto \bm{W}\bm{x}$ from the original $D$-dimensional space into a subspace with dimensionality $d = C - 1$, where $C$ is the number of classes. The matrix $\bm{W}$ is chosen such that the distance between the classes is maximized and the spread of each class is minimized.

\begin{ltxfigure}
\centering
\subfigure[Single diffractive left]{
\includegraphics[width=.48\textwidth]{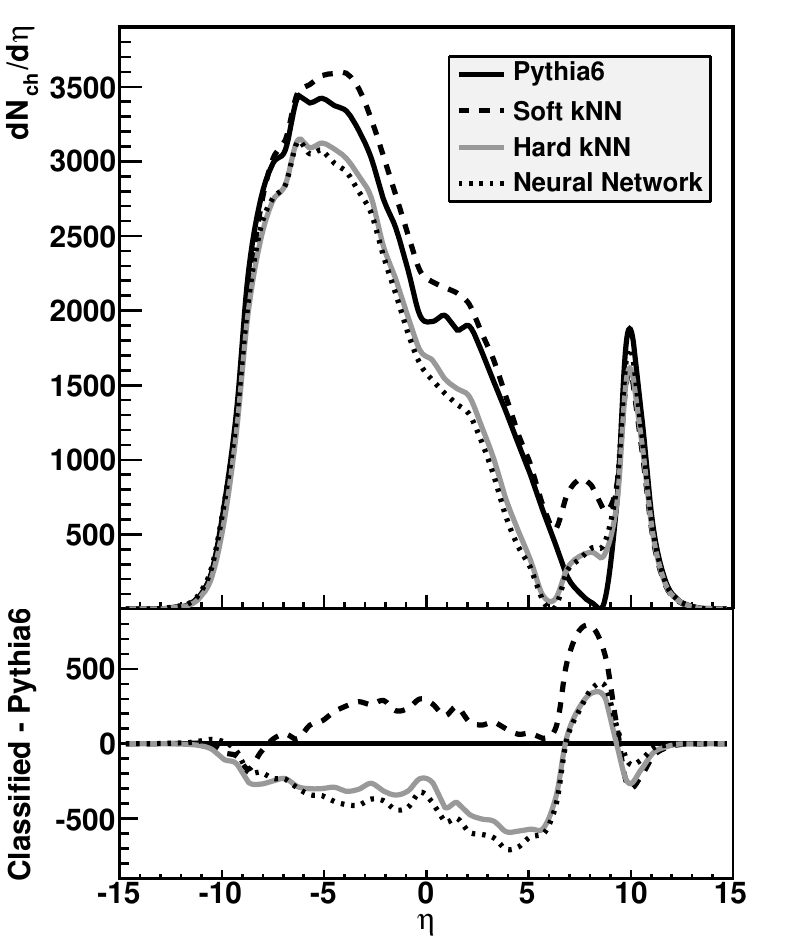}
\label{fig:sdl}}
\subfigure[Double diffractive]{
\includegraphics[width=.48\textwidth]{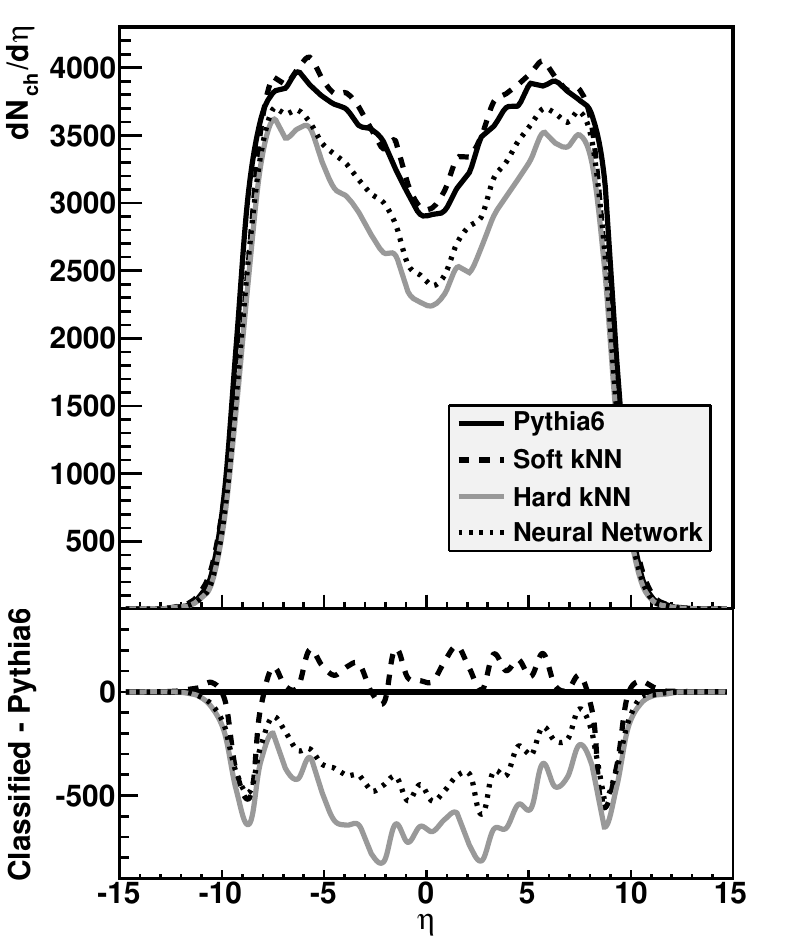}
\label{fig:dd}}
\caption{Charged particle multiplicity distributions for diffractive events when event categories are determined using different classification schemes. The distribution for the right side single diffraction is essentially a mirror image of the left side distribution shown here. The plots allow comparison between correctly labeled data ({\sc Pythia6}), soft classification (soft $k$NN) and hard classification (hard $k$NN and neural network). At central rapidities, hard classification underestimates all the diffractive contributions while soft classification is able to better reproduce the correct distributions. The accuracy of all the algorithms is impaired at large $|\eta|$, where information from only the ZDC detector is available.}
\label{fig:mult}
\end{ltxfigure}

\section{Soft Classification of Diffraction}

To study the feasibility of soft classification for distinguishing between the different diffractive classes, we generated a sample of $\sqrt{s} = 7$ TeV minimum bias events using {\sc Pythia6} with the D6T tune \cite{field2008}. The sample contained SDL, SDR, DD and ND events in ratios determined by the MC tune. Starting from this generator level information, we calculated energy deposits and charged particle multiplicities registered by the IP5 detectors at the LHC based on their geometric acceptances. By dividing the CMS central tracker into 3 $\eta$ bins and T1 and T2 on both sides into 2 bins, multiplicity was recorded in 11 $\eta$ bins. The same amount of bins was also used for energy deposits corresponding to division of the central calorimeters into 3 bins, HF on both sides into 2 bins and a single bin each for CASTOR and the Zero Degree Calorimeters (ZDC) on both sides of the interaction point. In the case of the ZDC, only the energy of neutral particles was recorded. No thresholds or other detector effects were included in the simulations. By computing also the scalar sum of $p_{\mathrm{T}}$ and the invariant mass of charged particles within $|\eta|<2.5$, each event was represented by 24-dimensional data vector $\bm{x}$.

The MC sample was divided into training, validation and test sets each containing 50000 events followed by a normalization with the mapping $x_i \mapsto \log(x_i + 1)$. After further normalization for mean and variance, the dimensionality of the events was reduced to 3 using LDA. The optimal value of the parameter $k$ for this data was found based on maximization of efficiency on the validation set. The $k$NN algorithm was then used to perform both soft and hard classification of the test set. As an additional benchmark, we also trained an MLP neural network \cite{alpaydin2010} with 10 hidden nodes on a single hidden layer to perform hard classification of the same test set.

The classification results were then used to reconstruct the multiplicity distributions of the different event types. The obtained diffractive distributions shown in Figure~\ref{fig:mult} indicate that soft classification is able to better reproduce the correct distributions than hard classification. Note also that both hard classification algorithms produce very similar outputs while the results of soft classification are qualitatively different from this. Similar results were also obtained for the $\sum p_{\mathrm{T}}$ distribution. We also observed that the relative event rates estimated using the soft $k$NN algorithm are very accurate (see Table~\ref{tab:rates}) and clearly better than the ones given by the hard methods.

\begin{table}
\caption{Relative event rates and their deviations from {\sc Pythia6} with the different classification schemes. Soft $k$NN is able to estimate the rates with a very high accuracy while both hard classification algorithms overestimate the non-diffractive contribution and underestimate all the diffractive classes.}
\begin{tabular}{lllll}
\hline
& \tablehead{1}{l}{c}{ND} & \tablehead{1}{l}{c}{DD} & \tablehead{1}{l}{c}{SDR} & \tablehead{1}{l}{c}{SDL} \\
\hline
{\sc Pythia6}  & 67.84  &  13.00  &  9.72  &  9.44 \\
Soft $k$NN & 67.66 (-0.18)  &  13.07 (+0.07)  &  9.78 (+0.06) &  9.48 (+0.04) \\
Hard $k$NN & 70.13 (+2.29) &  11.67 (-1.33) &  9.52 (-0.20) &  8.67 (-0.77)\\
Neural network &        69.67 (+1.83)  &  12.15 (-0.85) &  8.97 (-0.75)  &  9.20 (-0.24) \\
\hline
\end{tabular}
\label{tab:rates}
\end{table}

\section{Conclusions}

We propose a probabilistic multivariate approach called soft classification for identification and classification of diffraction. The results obtained using the soft $k$NN algorithm on a generator level MC sample show that the approach accurately reproduces physical observables. Soft classification could hence serve as an alternative to the rapidity gap method. The main drawback of the approach is its dependency on the selection of the training set which makes the classification MC dependent. The severity of this dependence is a subject of an ongoing study, the preliminary results of which suggest that soft classification is more robust against a misspecified training set than the hard methods. In some cases, it might also be possible to use data-driven methods for constructing the training set. The natural next step of the study is to employ detector level MC and eventually perform a full physics analysis using real data.


\begin{theacknowledgments}
The authors are grateful to the Academy of Finland for financial support and to Hannes Jung, Valery Khoze, Antonio Vilela Pereira and Tapani Raiko for valuable discussions and help with technical matters.
\end{theacknowledgments}



\bibliographystyle{aipproc}   

\bibliography{diff2010ref}

\IfFileExists{\jobname.bbl}{}
 {\typeout{}
  \typeout{******************************************}
  \typeout{** Please run "bibtex \jobname" to optain}
  \typeout{** the bibliography and then re-run LaTeX}
  \typeout{** twice to fix the references!}
  \typeout{******************************************}
  \typeout{}
 }

\end{document}